\documentclass[preprint,12pt]{aastex}
\usepackage{epsfig}
\usepackage{emulateapj5}
\usepackage{apjfonts}

\newcommand{\chandra}{{\it Chandra}}

\newcommand{\rosat}{{\it ROSAT}}

\newcommand{\xmm}{{\it XMM-Newton}}

\newcommand{\lum}{\thinspace\hbox{$\hbox{erg}\thinspace\hbox{s}^{-1}$}}

\newcommand{\sss}{XMMU\,J005510.7-373855}

\begin{document}

\def\spose#1{\hbox to 0pt{#1\hss}}
\def\laeq{\mathrel{\spose{\lower 3pt\hbox{$\mathchar"218$}}
     \raise 2.0pt\hbox{$\mathchar"13C$}}}
\def\gaeq{\mathrel{\spose{\lower 3pt\hbox{$\mathchar"218$}}
     \raise 2.0pt\hbox{$\mathchar"13E$}}}

\title{A Luminous Recurrent Supersoft X-ray Source in NGC 300}

\author{A.K.H.~Kong\altaffilmark{1} and R.~Di\,Stefano\altaffilmark{1,2}}
\altaffiltext{1}{Harvard-Smithsonian Center for Astrophysics, 60
Garden Street, Cambridge, MA 02138; akong@cfa.harvard.edu}
\altaffiltext{2}{Department of Physics and Astronomy, Tufts
University, Medford, MA 02155}

\begin{abstract}
We report the results of \xmm\ observations for an especially luminous
supersoft X-ray source (SSS) with bolometric luminosity of $10^{39}$
\lum\ in
the spiral galaxy NGC 300. The source was detected as a SSS in 1992 and
disappeared in subsequent X-ray observations. 
The source was active again during recent \xmm\ observations. 
It appeared to be very soft ($kT\sim 60$ eV) and very luminous ($\sim
10^{38}-10^{39}$ \lum). The two \xmm\
observations also reveal that the source went from a ``high'' state to
a ``low'' state in 6 days. We also found a 5.4-hr periodicity during
the ``low'' state. We consider white dwarf, black hole,
and neutron star models to explain the nature of the source.

\end{abstract}


\section{Introduction}

Supersoft X-ray sources (SSSs) form a distinct class of objects, first
established through 
\rosat\ observations. The hallmarks of SSSs are very soft X-ray emission
($kT$ typically $<100$ eV) and bolometric luminosities of $10^{36}-10^{38}$
\lum. The advent of \chandra\ and \xmm\ provides good opportunities to
detect and study 
SSSs in nearby galaxies. Luminous ($10^{38-40}$\lum) SSSs have been
found in several nearby galaxies including M31 (Kong et al. 2002;
Di\,Stefano et al. 2002a), 
M81 (Swartz et al. 2002), M101 (Pence et al. 2001; Mukai et al. 2003;
Di\,Stefano \& Kong 2003 [DK03]),
NGC4697 (Sarazin, Irwin \& Bregman 2001; DK03),
M51 (DK03), and M83 (Soria \& Wu 2003; DK03). 

NGC 300 is an SA(s)d galaxy, seen near face-on (inclination angle
46$^{\circ}$; Tully 1988) at a distance of $2.0\pm0.1$ Mpc (Freedman
et al. 2001). The galaxy has been observed by
\rosat\ (Read et al. 1997; Read \& Pietsch 2001) and \xmm\ (Soria \&
Kong 2003). 

In this Letter, we report the reappearance of a luminous SSS in NGC
300 as observed with \xmm.

\section{Observations and Data Reduction}

\subsection{\rosat}

NGC 300 was observed by \rosat\ five times from 1991 to 1997. A
detailed analysis of the \rosat\ data was done by Read \& Pietsch
(2001). Briefly, the datasets consist of 2 Position Sensitive
Proportional Counter (PSPC) and 3 High Resolution Imager (HRI)
pointings. The exposures range from $\sim 9$ ks to $\sim 37$ ks. The
luminous supersoft source was only detected in a 37 ks PSPC observation
taken on 1992 May and June (see Read \& Pietsch 2001). Spectral
analysis was also done by Read et al. (1997). We have
reanalyzed the PSPC spectrum taken on 1992 May/June, and used other data to set upper
limits on the luminosity for the long-term lightcurve.

We extracted the source spectrum from a $30''$ circular region, while
background was from an annulus region ($45''$ and $60''$ radii)
centered on the source. 
The spectrum was grouped into at least 20 counts per spectral bin
to allow $\chi^2$ statistics to be used.

\subsection{\xmm}

The \xmm\ instrument modes were full-frame, medium filter for the three European Photon
Imaging Cameras (EPIC).
The first observation was taken on 2000 December 26 for 37 ks and the
second observation was on 2001 January 1 for about 47 ks. 
After rejecting intervals with a high
background level, we considered a good time interval of $\sim 28$ ks
and $\sim 40$ ks for the first and second observation, respectively. 
Data were reduced and analyzed
with the \xmm\ SAS package v5.3.3.
%\footnote{http://xmm.vilspa.esa.es/external/xmm\_sw\_cal/sas\_frame.shtml}.

We used here the MOS images to determine the position of the source
because the spatial resolution of MOS detector is slightly better than
that of pn. The source is located at 
R.A.=00h55m10s.7, Dec.=$-37^{\circ}38'55''$ (J2000), $\sim 4'$ (2.4 kpc) from the
galactic center; the derived positions from the two \xmm\ observations agree
with each other. This position is about $5''$ off from previous \rosat\ PSPC
observation, and is consistent with the positional error ($7.3''$)
quoted by Read \& Pietsch (2001). 

Source spectra and lightcurves of \sss\ were extracted with the SAS task
{\sc xmmselect}. Source-free regions were used for background to avoid
the chip boundary and a nearby faint source.
In order to allow $\chi^2$ statistics to be used, all
the spectra were binned such that there are at least 20 counts per
spectral bin. Response matrices were created by {\sc rmfgen} and {\sc arfgen}. 

\section{Analyses and Results}

%\subsection{Spectral Analysis}

Spectral analysis was performed by making use of XSPEC v11.2.
%\footnote{http://heasarc.gsfc.nasa.gov/docs/xanadu/xspec/index.html}.
Table 1 shows the best-fitting spectral parameters for the three
\rosat\ and \xmm\ observations.

For the \rosat\ observation, the spectrum (see Figure 1) can be fit with a blackbody
model with $N_H=1.1\times10^{21}$ cm$^{-2}$ and $kT=48.7$ eV; the
0.2--2 keV luminosity is $10^{39}$ \lum. We note that Read et
al. (1997) fit the spectrum with a thermal bremsstrahlung model
with temperature of a 0.1 keV and we confirmed that it is also an
acceptable model.  Using the thermal bremsstrahlung model, the 
luminosity becomes $3.8\times10^{38}$ \lum\ [the luminosity quoted by
Read et al. (1997) is $1.6\times10^{37}$ \lum, which is an absorbed
luminosity corrected for
absorption in our own Galaxy], a factor of 2.5 lower than 
the blackbody model. 
While we cannot distinguish between the two models
statistically, we prefer the blackbody model as subsequent \xmm\
observations confirm the supersoft nature of the source (see
below). Also, the \xmm\ observations would have detected the high
energy photons associated with the thermal bremsstrahlung model, had
the model been correct. Finally, the \xmm\ observations provide more
photons than \rosat.

We fit the pn, MOS1 and MOS2 data simultaneously with
several single-component models with interstellar absorption
(including absorbed power-law, thermal bremsstrahlung, blackbody and
Raymond-Smith); only the blackbody model provides an acceptable
fit. The blackbody temperature of both observations ranges between 57
eV and 67 eV, while the $N_H$ varies from $1.5\times10^{21}$ cm$^{-2}$
in the first observation to $5.3\times10^{20}$ cm$^{-2}$ in the second
observation. The 0.2--2 keV luminosity also drops from
$9\times10^{38}$ \lum\ to $1.2\times10^{38}$ \lum, indicating that the
source is a variable on timescale of days. The spectra of the
two observations are shown in Figure 1.

%\subsection{Temporal Analysis}

%\subsubsection{Long-term Variability}
The combined \rosat\ and \xmm\
long-term lightcurve of the SSS is shown in Figure 2, which is constructed
from a series of \rosat\ and \xmm\ pointings. The source was below the
detection limit of
other \rosat\ observations. In these cases, we estimated the $3\sigma$
limits, assuming a blackbody model
with mean $N_H$ ($10^{21}$ cm$^{-2}$) and $kT$ (58 eV). The value of 
$N_H$ plays an important role in estimating the luminosity. For instance,
if we lower the $N_H$ to $5.3\times10^{20}$ cm$^{-2}$, as found in the
second \xmm\
observation, the upper limits decrease by a
factor of $\sim 3$. In other words, the source varies by as much as
factor of 30 between the ``low'' state and the ``high'' state spanning
8.5 years.

%\subsubsection{Short-term Variability}

To search for short-term variability, we computed the Lomb-Scargle
periodogram (LSP; Lomb 1976; Scargle 1982), a modification of the
discrete Fourier transform which is generalized to the case of uneven
spacing. In each observation, we extracted the combined background
subtracted lightcurve
from pn and MOSs to increase the signal-to-noise. Individual
lightcurves from each of
the three detectors were also used to verify the result. By
applying the LSP to the combined lightcurve, we found that there is a
sharp peak at 5.4 hr in
the ``low'' state observation (see Figure 3). Independent checks from the pn and
MOS data also confirmed the periodicity. We determined the 99.9\% confidence
level by generating Gausiaan noise datasets with the same time
intervals and variance as the true data, and then performed the LSP
analysis on the resulting datasets (see Kong, Charles \& Kuulkers
1998). The peak power in each periodogram
(which must be purely due to noise) was then recorded. This was
repeated 10000 times to obtain good statistics. The peak at 5.4 hr is well
above the 99.9\% confidence level. The folded lightcurve of the ``low'' state data in 5.4 hr
is also shown in Figure 3.

Similar analysis was also carried out in the ``high'' state data, but
there is no significant peak in the LSP. The folded lightcurve shows
no obvious periodic variability at 5.4 hr (see Figure 3). It is not
clear if it is due to geometric effect but the exposure time of the ``high'' state 
covers less than 8 hr, corresponding to 1.4 cycles of the 5.4-hr
period in contrast to the 2.1 cycles in the ``low'' state. We
therefore performed 
Kolmogorov-Smirnov test to test
for variability and found significant variability at the 99.9\%
levels. We also search for variability of the hardness ratio but no
significant change was found in both observations.

%\subsection{Optical}

We also examined an Digitized Sky Survey image of NGC 300 to search for
possible optical counterpart. Within a $4''$ error circle of the \xmm\
position, there is no obvious optical counterpart with limiting $B$ and $R$
magnitudes of 21 and 18.1, respectively. However, the source is
surrounded by H{\sc ii} regions (Deharveng et al. 1988; Blair \& Long 1997), OB associations
(Pietrzynski et al. 2001) and supernova remnants (Blair \& Long
1997). The nearest H{\sc ii} region (source 136; Deharveng et al. 1988)
is located about $20''$ ($\sim 200$ pc) south-east of the X-ray
source. Within $50''$ ($\sim 500$ pc) of the SSS, we found an OB
association (AS99;
Pietrzynski et al. 2001), two supernova remnant candidates (N300-S21,
N300-N23; Blair \& Long 1997) and a H{\sc ii} region (N300-H20; Blair \&
Long 1997). 
We also looked for UV counterpart
from the Optical Monitor (OM) image (see Soria \& Kong 2003 for details of the OM
observations); no 
counterpart is found at the position of the SSS in both observations.

\section{Discussion}

It is clear from the spectral fits of all \rosat\ and \xmm\
observations that the \sss\ is very soft ($kT \sim 48-67$ eV) and is
 highly variable. Such a soft spectrum is consistent with SSSs seen in
our own Galaxy, the Magellanic Clouds, M31 (Greiner 2000 \footnote{see
http://www.aip.de/$\sim$jcg/sss/ssscat.html for updated catalog}) and
several nearby galaxies (see \S\,1). Since the Galactic and Magellanic
Cloud SSSs are conjectured
to be white dwarfs (WDs), it is reasonable to consider    
a WD model for the source in NGC 300. 
Since, however, the luminosity in the high
state may be too large to be consistent with the standard SSS WD model,
we also consider black hole (BH) and neutron star (NS) models. 

%\subsection{WD Models}

%In this model, the emission comes from a WD that burns accreting
The X-ray emission of the SSS could come from a WD that burns accreting
hydrogen in a quasi-steady manner. The maximum luminosity is the
Eddington luminosity for a $1.4M_{\odot}$ object. WD models therefore
seem to be ruled out by the high bolometric luminosities during 2 of
the observations. It is nevertheless worth noting that the WD model
temperatures are comparable to those we observed, and that
X-ray variability by a factor
of a few over times of days and months has been observed in SSSs.
(see e.g., Greiner and Di Stefano 2002 and references therein.)  
If the 5.4-hour variation detected during the X-ray low state corresponds 
to an orbital period, then this system may be similar
to 1E0035.4-7230, a Magellanic Cloud SSS that appears to have a
 $4$ hour period (Kahabka 1996). In such a short-period binary, it is likely that 
a radiation-driven wind plays an important role in the
binary evolution. (see, e.g., van Teeseling \& King 1998)
A WD model would perhaps be the most natural explanation for the
SSS in NGC 300, were it not for the high estimated bolometric
luminosities. 
Note that it may be difficult to use beaming to circumvent the
Eddington limit in these systems, since the inner region of the accretion
disk can also become a luminous emitter of SSS radiation 
(Popham \& Di\thinspace Stefano 1996).  

%\subsection{Black Hole Models}

SSS emission is expected from accreting BHs.
Modeling the accretion disk as a thin disk which is optically thick,
we can derive a relationship between the minimum mass of the accretor,
and the observed temperature and luminosity (Frank, King, \& Raine 2002). Using $kT=60$ eV and
$L_X=10^{38}$\lum\ ($10^{39}$\lum), we find that the accretor mass is
greater than $\approx 890 M_{\odot}$ ($2800 M_{\odot}$). Although the
fact that the observed luminosity is $10^2-10^3$ times smaller than
the Eddington luminosity may call the validity of the disk model into
question, this calculation suggests that the accretor is an
intermediate-mass BH. If the accretor is a BH, we may use $L_{obs}=0.1
\dot m c^2 (\eta/0.1)$, where $\dot m$ is the accretion rate and
$\eta$, the efficiency factor, is likely to be close to 0.1. This
yields $\dot m\approx 1.8\times10^{-8} M_{\odot}$/yr
($1.8\times10^{-7} M_{\odot}$/yr) for $L_{obs}=10^{38}$\lum\
($L_{obs}=10^{39}$\lum). This would be consistent with an
irradiation-driven wind from a low-mass donor. Measurement of the
orbital period could provide supporting evidence. Indeed, if the donor
fills its Roche lobe, then $P_{orb}\approx(8.9~\mathrm{hrs})(M_d/M_{\odot})$,
where $M_d$ is the mass of the donor star. If $P_{orb}=5.4$ hrs, then
$M_d\approx0.61 M_{\odot}$.

%\subsection{Neutron Star Models}

It is of course possible that the observed emission emanates from an
accreting NS. In this case the SSS emission would presumably emanate
from a photosphere which is much larger than the neutron star itself.
The photospheric radius would be different in each observation:
$3.7 \times 10^9$ cm during the {\it ROSAT} observation,
$2.6 \times 10^9$ cm during the \xmm\  high-state observation, 
and $6.8 \times 10^8$ cm during the \xmm\ low-state observation,
corresponding to the different temperatures and luminosities.

At this point, neutron star models seem conceptually unattractive
because there is no obvious explanation for why the photosphere should
achieve these relatively large sizes, and no way to
relate them to the system's fundamental physical parameters.
In addition, just as in the WD models, the luminosity
appears to be super-Eddington during two of the observations, 
unless the neutron star has a mass as large as $3-4\, M_\odot.$
In this case, if the efficiency of turning accretion energy into X-rays is the 
same during all $3$ observations,  
then the accretion rate must have changed by an order of magnitude,
approaching $2\times 10^{-8} M_\odot$ yr$^{-1}$ during the high state.
(see Di\thinspace Stefano et al. 2002b for a similar model of the
X-ray source in the globular cluster Bo 375.)  
If, on the other hand, the mass of the neutron star is close to $1.4\, M_\odot,$
then other ways around the Eddington limit must be found; e.g.,
the energy could be beamed into a smaller solid
angle during the $2$ high-state observations.
In this case, the accretion rate could be significantly lower.  

\section{Summary}

We have studied a recurrent luminous SSS in NGC300 with \xmm. The
source was seen by \rosat\ in 1992 and fell below the detection limit
in subsequent \rosat\ observations. It reappeared in recent \xmm\
observations with bolometric luminosities between $10^{38}$ and
$10^{39}$ \lum. During the ``low'' state, the source showed a 5.4-hr
periodicity. If we consider the periodicity is due to the orbital
period, then the X-ray emission can be explained by WD, BH, and NS
models. However, WD and NS models appear to be unlikely due to the high X-ray
luminosity during the ``high'' state. Further repeated X-ray
observations of the SSS in different states may permit discrimination among these
models.

\begin{acknowledgements}
This work was supported by NASA under an LTSA grant,
NAG5-10705. A.K.H.K. acknowledges support from the Croucher Foundation.
%This work is based on observations obtained with \xmm, an ESA mission
%with
%instruments and contributions directly funded by ESA member states and
%the US (NASA). 
\end{acknowledgements}

\newpage

\begin{deluxetable}{cccccc}
\tabletypesize{\small}
\tablecaption{Best-fitting Spectral Parameters}
\tablewidth{0pt}
\tablehead{
\colhead{Date} & \colhead{$N_H$} & \colhead{$kT$} &
\colhead{$L_X$}\tablenotemark{a} & \colhead{$L_{bol}$}\tablenotemark{b}
& \colhead{$\chi^2_{\nu}$/d.o.f}\\
 &  ($\times10^{21}$ cm$^{-2})$ &(eV) & & &\\
}
\startdata
1992 May 26  & $1.08^{+0.31}_{-0.31}$ & $48.7^{+9.1}_{-9.1}$
& 10 & 20& 0.74/11\\
2000 Dec 26 & $1.50^{+0.32}_{-0.20}$ &
$56.8^{+2.2}_{-3.3}$ & 9 & 15 & 1.07/39\\
2001 Jan 1 & $0.53^{+0.20}_{-0.07}$ &
$66.5^{+2.4}_{-3.6}$ & 1.1 & 1.9 & 1.15/48\\ 
\enddata
\tablecomments{All quoted uncertainties are 1$\sigma$.}
\tablenotetext{a}{0.2--2 keV luminosity ($\times 10^{38}$\lum),
assuming a distance of 2 Mpc}
\tablenotetext{b}{Bolometric luminosity ($\times 10^{38}$\lum)}

\end{deluxetable}

\newpage

\begin{figure}
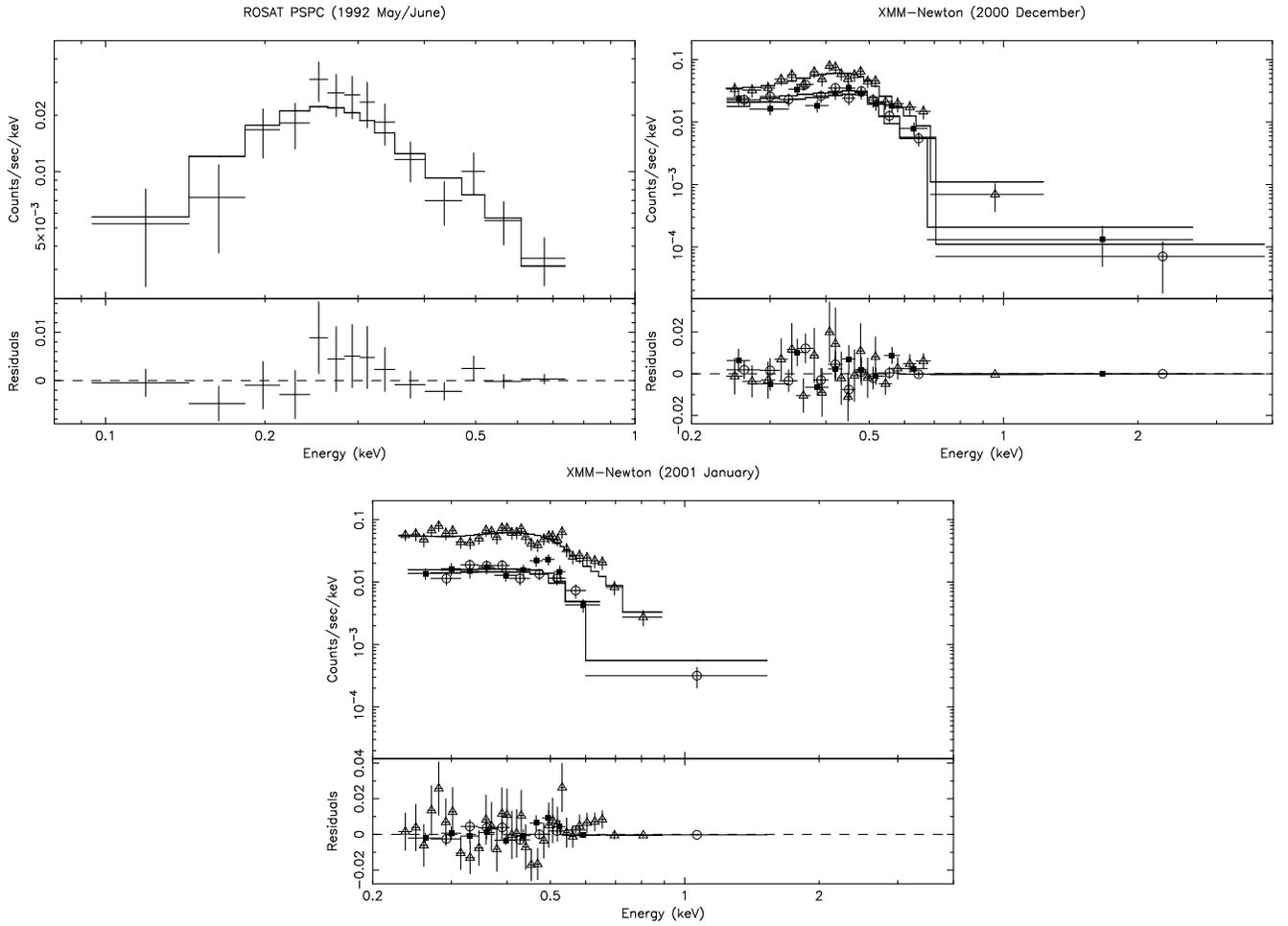

\begin{center}
{\rotatebox{-90}{\psfig{file=f1a.ps,width=2.5in}}}
{\rotatebox{-90}{\psfig{file=f1b.ps,width=2.5in}}}
{\rotatebox{-90}{\psfig{file=f1c.ps,width=2.5in}}}
\end{center}
\caption{Energy spectra of \sss\ (see Table 1 for spectral parameters).
From left to right: \rosat\ PSPC spectrum on 1992 May/June (high state), \xmm\ spectrum on
2000 December 26 (high state) and \xmm\ spectrum on 2001 January 1
(low state). For the \xmm\ spectra, pn, MOS1, and MOS2 data are marked
as triangles, solid squares and circles.}
\end{figure}

\begin{figure}
\begin{center}
\psfig{file=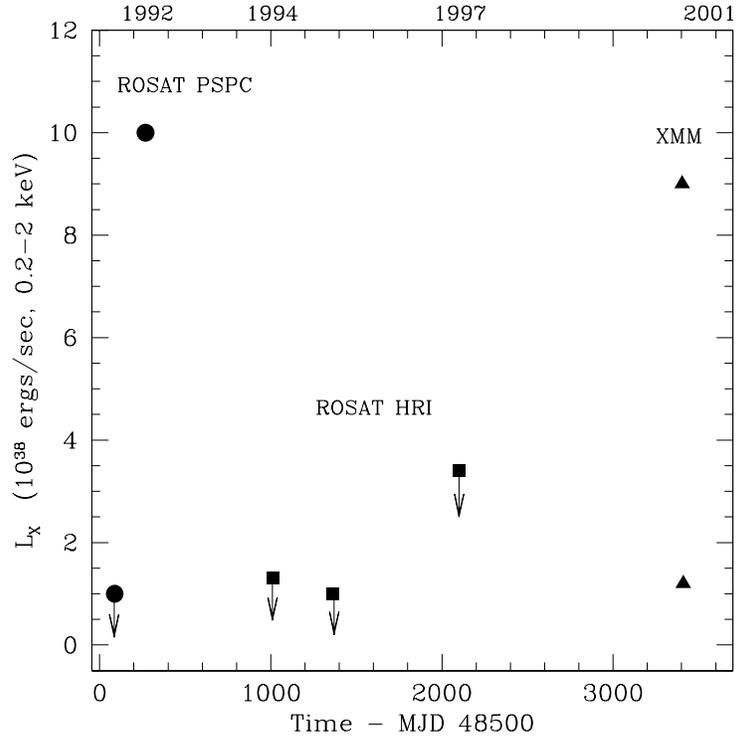,width=4in}
\end{center}
\caption{Long-term lightcurve of the supersoft ULX from 1991 to 2001
(circles: \rosat\ PSPC; squares: \rosat\ HRI; triangles: \xmm). The 0.2--2 keV
luminosity is from spectral fits in Table 1, while the $3\sigma$ upper
limits are assuming a blackbody model with $N_H=10^{21}$ cm$^{-2}$ and
$kT=58 eV$.}
\end{figure}

\begin{figure}
\begin{center}
\psfig{file=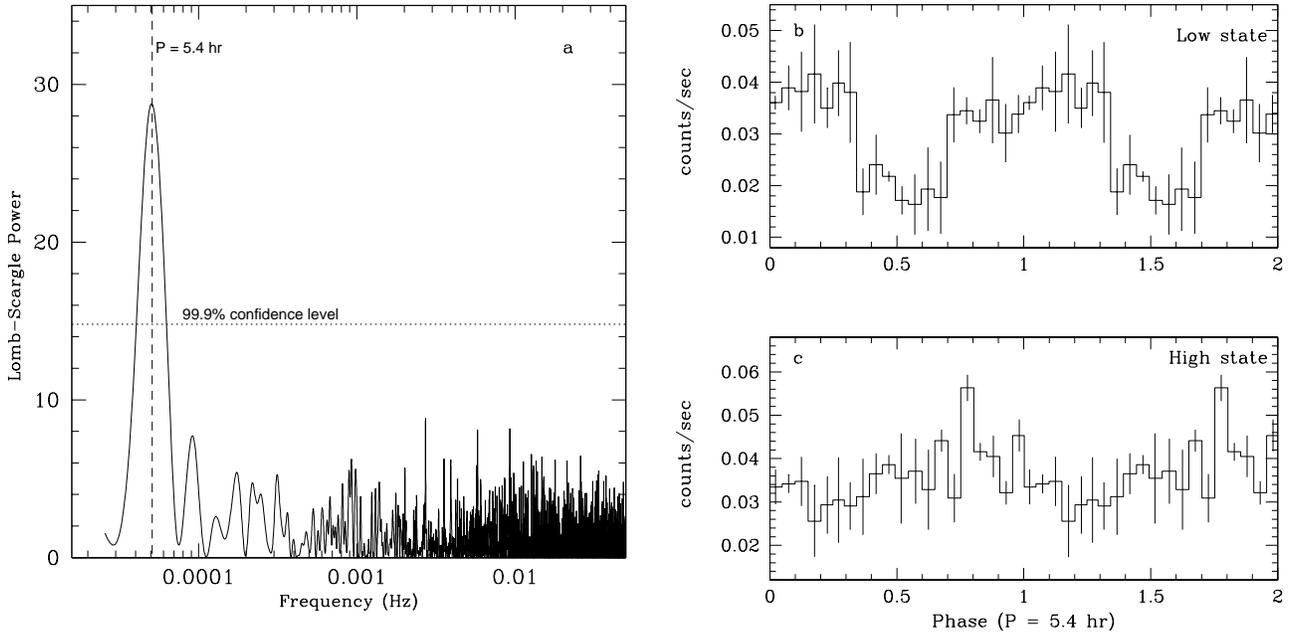,width=7in}
\end{center}
\caption{a) Lomb-Scargle periodogram of \sss\ as obtained by \xmm\ on
2001 January. The horizontal dotted line is the 99.9\% confidence
level; b) Folded lightcurve of the ``low'' state (2001 January) data
on a period of 5.4 hr; c) Folded lightcurve of the ``high'' state (2000 December) data
on a period of 5.4 hr. $T_0$ of both lightcurves are set at the time of the first
data point.}
\end{figure}

\end{document}